\documentclass[pss]{wiley2sp}
\usepackage{amsmath}
\usepackage{color}
\usepackage{graphicx}
\usepackage{cancel}
\usepackage{amsfonts}

\newcommand{\ud}{\,\mathrm{d}}
\DeclareMathOperator{\e}{e}

\begin{document}

\title{Dynamics of the polarization of a pinned domain wall in a magnetic nanowire}

\titlerunning{Dynamics of a pinned domain wall}

\author{%
  N. Sedlmayr\textsuperscript{\Ast,\textsf{\bfseries 1}},
  V. K. Dugaev\textsuperscript{\textsf{\bfseries 2,3}},
  J. Berakdar\textsuperscript{\textsf{\bfseries 4}}}

\authorrunning{N. Sedlmayr et al.}

\mail{e-mail
  \textsf{sedlmayr@physik.uni-kl.de}, Phone: +49-631-205 3159}

\institute{%
  \textsuperscript{1}\,
  Department of Physics and Research Center OPTIMAS, University of Kaiserslautern, 67663 Kaiserslautern, Germany\\
  \textsuperscript{2}\,
  Department of Physics, Rzesz\'ow University of Technology,
  Al.~Powsta\'nc\'ow Warszawy 6, 35-959 Rzesz\'ow, Poland\\
  \textsuperscript{3}\,
  Department of Physics and CFIF, Instituto Superior T\'ecnico,
  TU Lisbon, Av.~Rovisco Pais, 1049-001 Lisbon, Portugal\\
  \textsuperscript{4}\,
  Martin-Luther-Universit\"at Halle-Wittenberg, Heinrich-Damerow-Str. 4,
  06120 Halle, Germany}



\abstract{
We consider the dynamics of polarization of a single domain wall
in a magnetic nanowire, which is strongly pinned by impurities. In this case the equation of motion
for the polarization parameter does not include any other dynamical variables
and is nonlinear due to magnetic anisotropy.  We calculated numerically the magnetization
dynamics for different choices of parameters under short current pulses inducing
polarization switching. Our results show that the switching is most effective for
very rapid current pulses. Damping also enhances the switching probability.}

\maketitle

\section{Introduction}

Since the possibility of moving magnetic domain walls (DWs) in nanowires due
to electric currents was
realized, the behaviour of DWs in wires subject to a variety of pulses, currents, fields and pinning
forces has been extensively studied \cite{Marrows2005,Thiaville2004,Berger1996,Braun1996,Sedlmayr2009,Sedlmayr2011a}.
Electrons passing a DW transfer momentum and spin with the
DW, which results in a modification of the resistivity of the wire and the motion of the
DW \cite{Ara'ujo2006,Dugaev2006,Tserkovnyak2008}.
In simple scenarios the equations governing the DW motion can be set up and
solved \cite{Landau2002,Malozemoff1979,Schryer1974,Bouzidi1990,Takagi1996,Slonczewski1996,Gilbert2004,Tatara2004}.
The ability to move domain walls around raises possible technological applications, in particular
the possibility of constructing a solid state memory with magnetic domains playing the role of
bits \cite{Parkin2008,Thomas2010}. In these works the DW dynamics was mostly investigated
for the translational motion along the wire.

Then in an experiment performed on a Permalloy nanowire it was found that the polarity of a transverse DW
can be controlled with current pulses \cite{Vanhaverbeke2008,Klaui2008}.
Additionally, it was shown that the DW polarity is mainly determined by the direction of the current.
Since then attention has also been attracted to the DW dynamics of translationally non-invariant
systems \cite{Martinez2011,Tretiakov2012}.

In this work we concentrate on the dynamics of a strongly pinned DW, which is controlled by
short current pulses. In this case the possible motion along the wire is limited to a very
short distance.
We use a model with a single DW pinned to the potential just like a
particle in an oscillatory potential. This model allows us to study in a simple way how
the rotation of the DW polarization is related to the oscillatory longitudinal DW motion.

\section{Model}

Let us start from the Lagrangian density describing the spin dynamics of a one dimensional
domain wall (DW) \cite{Braun1996,Bouzidi1990,Takagi1996,Tatara2004}
\begin{eqnarray}
\label{Lagrangian}
\mathcal{L}&=&\hbar S\dot{\phi }(\cos \theta -1)
-\frac{S^2}2\Big\{ J[(\nabla \theta )^2+\sin ^2\theta (\nabla \phi)^2]
\nonumber \\&&\qquad+\sin ^2\theta \, (K+K_\bot \sin ^2\phi )\Big\}-V_{pin}\,,
\end{eqnarray}
where the angles $\theta =\theta (x,t)$ and $\phi (x,t)$ determine the spin
orientation in the DW:
\begin{equation}
\hat{n}(x,t)=\begin{pmatrix}\cos[\phi(x,t)]\cos[\theta (x,t)]\\
\sin[\phi(x,t)]\cos[\theta (x,t)]\\\sin[\theta (x,t)]\end{pmatrix}\,.
\end{equation}
$S$ is the localized spin, $J$ the exchange coupling between spins and $K$ and $K_{\perp}$ the magnetic anisotropy. We have introduced a pinning field $V_{pin}\equiv SV_0a^2/2$ where $a$ is the position of the DW. In the following we set $\hbar=1$

\begin{figure}
\includegraphics[width=0.95\columnwidth]{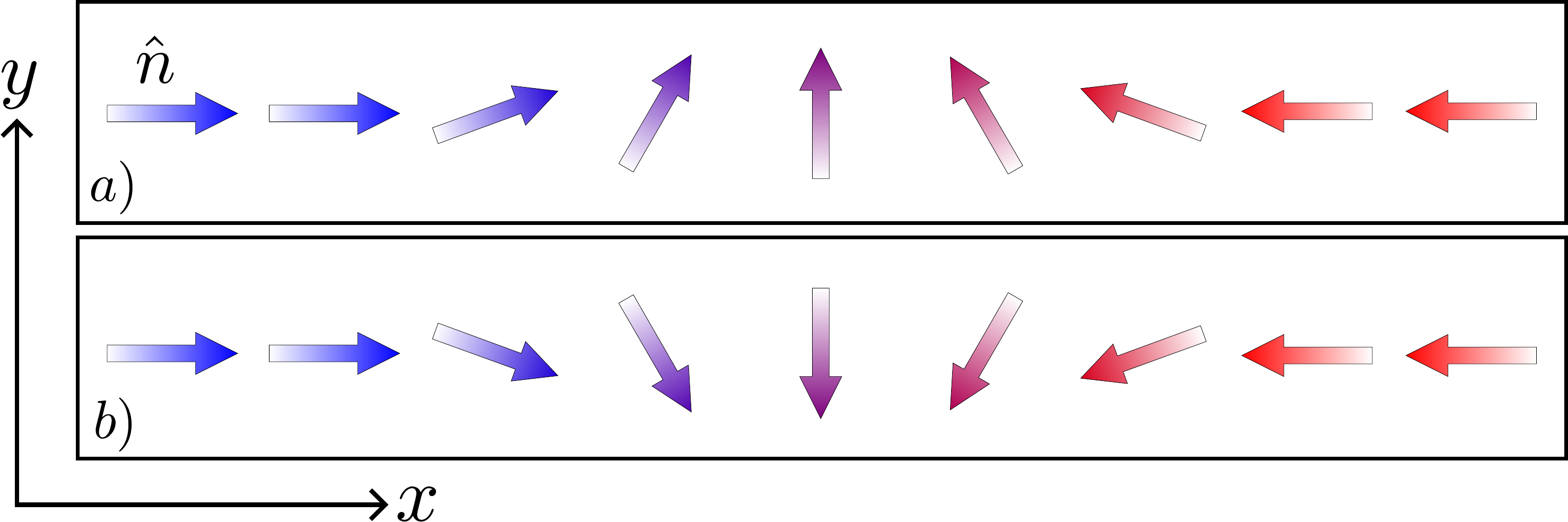}
\caption[DW Schematic]{A schematic of the two stable in plane DW polarizations, a) and b) in the wire. The polarization dynamics flips the system between these two orientations.}
\label{schematic}
\end{figure}

Assuming that a single DW is strongly pinned at the point $x=0$,
we can use the following parametrization with two functions $a(t)$ and $f(t)$
\begin{eqnarray}
\label{ansatz}
\theta (x,t)&=&\theta _0\left(x-a(t)\right),\nonumber
\\
\phi (x,t)&=&\psi _0\left(x-a(t)\right) f(t),
\end{eqnarray}
where we assume that $\theta _0(x)$ and $\psi _0(x)$ are known
functions describing the DW shape.
The time dependent functions $a(t)$ and $f(t)$ describe the fluctuation of the DW
location and the polarization dynamics around the pinning site $x=0$, respectively. We are ineterested in the dynamics of the polarization between two stable in plane configurations shown schematically in Fig.~\ref{schematic}.

We use the profile of a static DW\cite{Thiaville2002,Thiaville2006} $\theta _0(x)=\cos ^{-1} \tanh (x/\lambda )$.
As we are interested only in strongly pinned DWs it is sufficient to consider a constant DW width $\lambda$.
$\psi _0(x)$ is chosen to impose
a similar characteristic DW width $\lambda _\phi $ to the variation of angle $\phi $ within the wall.
If we assume $\lambda\ll\lambda_\phi$ then $\psi _0(x)\approx 1$.
The Lagrangian density Eq.~\eqref{Lagrangian} becomes
\begin{eqnarray}
\label{Lagrangian2}
\mathcal{L}
\simeq \hbar S\psi _0\, \dot{f}\, [\cos \theta _0-1]-\frac{SV_0a^2}2
\hskip1cm
\\
-\frac{S^2}2\Big\{ J\ \left( \nabla \theta _0\right) ^2
+\sin ^2\theta _0\, \big[ K+K_\bot \sin ^2\big( f\, \psi _0\big) \big] \Big\}.\nonumber
\end{eqnarray}
We obtain dynamical equations for $a(t)$ and $f(t)$ by variation of the Lagrangian
$L=\int dx\, \mathcal{L}$ over $a(t)$ and $f(t)$.
Noting that $\theta _0(x)$
is an odd function of $x$ we obtain
the coupled equations
\begin{eqnarray}
\label{9}
&&2\dot{f}=-V_0a\,,
\\
&&2\dot{a}=SK_\bot \int \frac{\sin (f\psi _0)\, \cos (f\psi _0)\, \psi _0\, dx}
{\cosh ^2(x/\lambda )}\,,
\end{eqnarray}
where we have used that for $\lambda _\phi \gg \lambda$
\begin{eqnarray}
\label{11}
-\int\ud x \psi _0\sin \theta _0\, \theta '_0&\approx&2\,,\textrm{ and}
\\
\int\ud x \sin ^2\theta _0(\nabla \psi _0)^2 &\approx& 0\,.
\end{eqnarray}

We can now add damping, $\alpha$, and torque, $T$, to the equations of
motion\cite{Tatara2004}:
\begin{eqnarray}
\label{fdot}
-2\left( \dot{f}+\frac{\alpha \dot{a}}{\lambda}\right) &=&V_0a+T_t\,\textrm{, and}
\\\label{adot}
-2\left( \dot{a}-\alpha \lambda \dot{f}\right) &=&SK_\bot \lambda \sin (2f)+T_l\,.
\end{eqnarray}
$T_{t(l)}$ is the transverse (longitudinal) torque.

To find the dynamical equation for $f(t)$ we
differentiate Eqs.~\eqref{fdot} and \eqref{adot} with respect to time resulting in
\begin{equation}
\label{17}
2\ddot{f}+\frac{2\alpha \ddot{a}}{\lambda }=-V_0\dot{a}-\dot{T}_t\,,
\end{equation}
and
\begin{equation}
\label{18}
\ddot{a}=\alpha \lambda \ddot{f}-SK_\bot \lambda \dot{f}\cos (2f) -\frac{\dot{T}_l}{2}\,.
\end{equation}
Then we find
\begin{eqnarray}
\label{eqn_motion_f}
&&(1+\alpha ^2)\ddot{f}
-\left(S\alpha K_\bot \cos (2f)-\frac{V_0\alpha \lambda }{2}\right) \dot{f}
\nonumber \\
&&-\frac{V_0SK_\bot \lambda \sin (2f)}{4}-\frac{\alpha \dot{T_l}}{2\lambda }
-\frac{V_0T_l}{4}+\frac{\dot{T}_t}{2}=0\,.
\end{eqnarray}
This is the equation of motion for the polarization parameter $f$ of the pinned DW.
Here the dynamics depends on the magnitude and the time derivative of longitudinal
and transverse components of the spin torque.\cite{Bocklage2009}
In the limiting case of strong pinning $V_0\lambda\gg SK_\perp$ and Eq.~\eqref{eqn_motion_f} could be further simplified.

\section{Results}

\begin{figure}
\includegraphics[width=0.95\columnwidth]{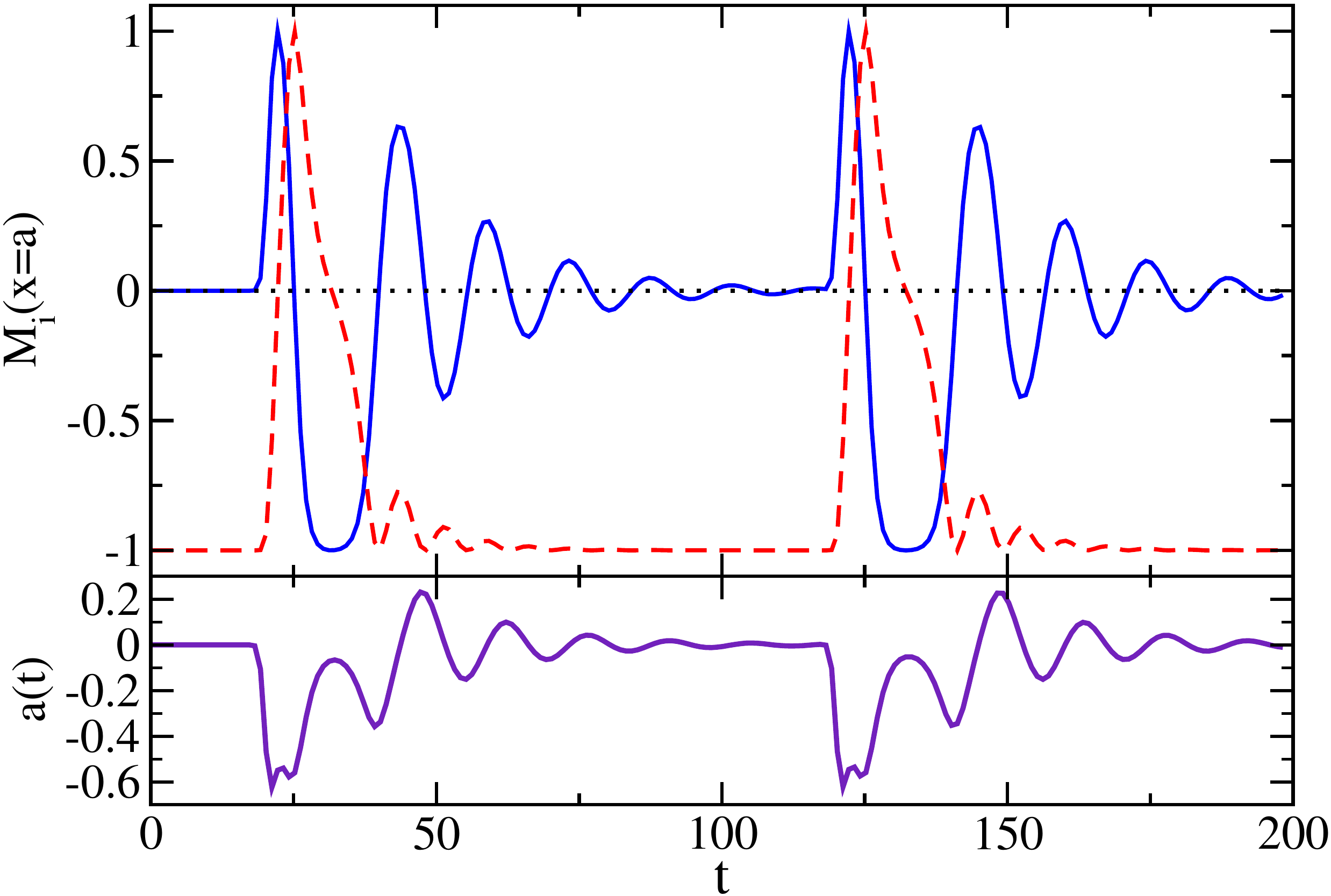}
\caption[DW Motion]{The DW motion for $T_t=\tilde{T}(t-20)+\tilde{T}(t-120)$ with $T_0=\beta=1$ and $\alpha=0.1$, see Eq.~\eqref{pulse}. Shown is the motion of the centre of the domain
wall $a(t)$ and the DW orientation at the centre: $M_i(x=a(t),t)$ for $i=x$ (blue, straight),
$i=y$ (red, dashed), and $i=z$ (black, dotted).}
\label{DW_Motion_Example1}
\end{figure}

We now solve Eq.~\eqref{eqn_motion_f} for particular scenarios. In the following numerical calculations we set $SK_\perp=0.2$ and $\lambda V_0=2$, which satisfies the strong pinning condition. Furthermore we only apply a transverse torque so that $T_l=0$. As there is no deformation of the domain wall the exchange strength itself does not appear as a parameter of the dynamics. One can then define a Gaussian shaped pulse:
\begin{equation}\label{pulse}
\tilde{T}(t-t_0)=T_0 \e^{-(t-t_0)^2\beta}\,,
\end{equation}
which starts the DW dynamics. We are interested in how changes on the torque strength and shape affect the polarization dynamics of the DW. The DW is always started in an equilibrium position with $f(t=0)=-\pi/2$.

\begin{figure}
\includegraphics[width=0.95\columnwidth]{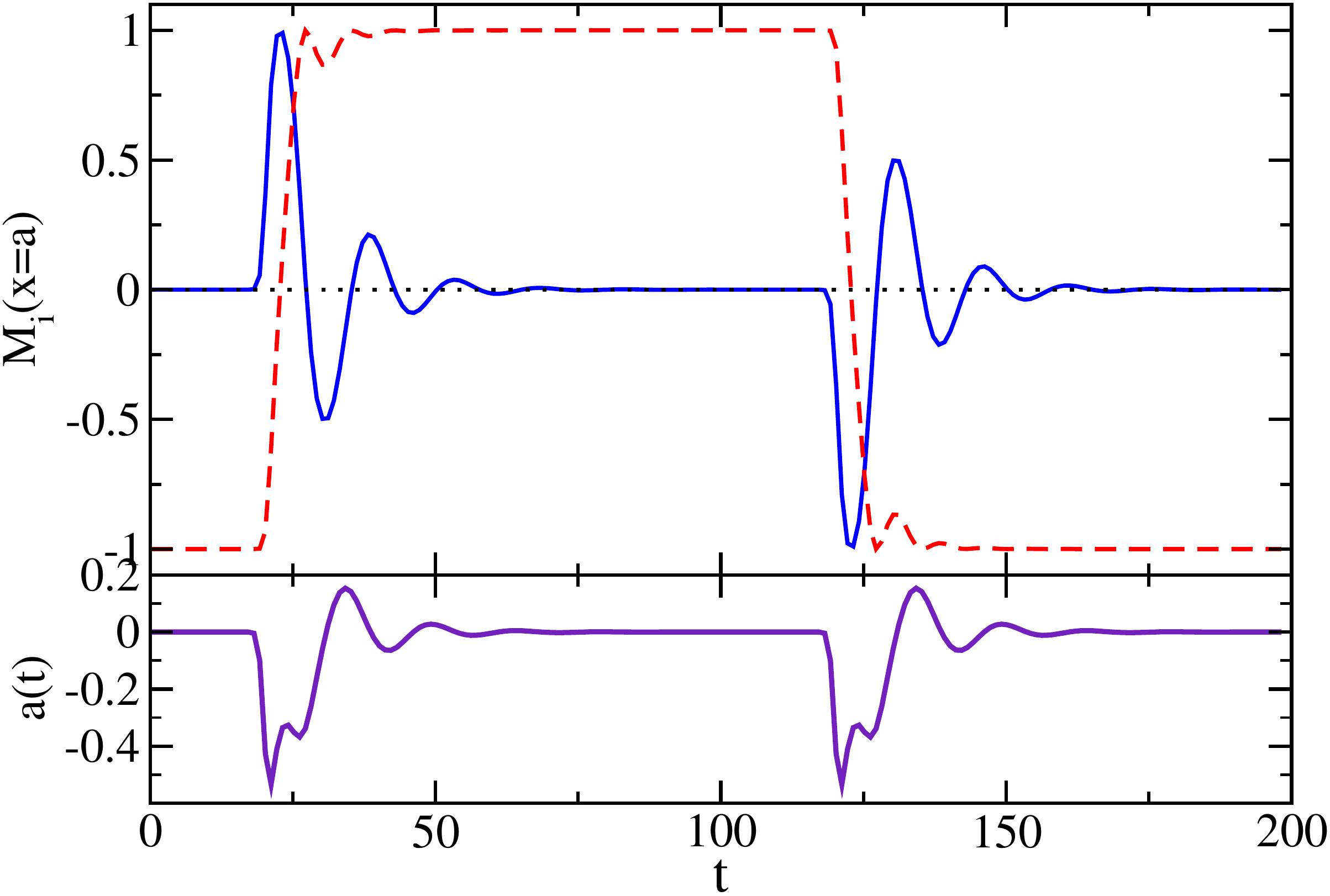}
\caption[DW Motion]{The DW motion for $T_t=\tilde{T}(t-20)+\tilde{T}(t-120)$ with $T_0=1$, $\beta=0.25$ and $\alpha=0.2$, see Eq.~\eqref{pulse}. Shown is the motion of the centre of the domain
wall $a(t)$ and the DW orientation at the centre: $M_i(x=a(t),t)$ for $i=x$ (blue, straight),
$i=y$ (red, dashed), and $i=z$ (black, dotted).}
\label{DW_Motion_Example2}
\end{figure}

Solving Eq.~\eqref{eqn_motion_f} for the case of two successive pulses we find the distribution of
magnetization orientation and the position of DW centre as a function of time under
current pulses of intermediate duration: $T_t=\tilde{T}(t-20)+\tilde{T}(t-120)$. The results are shown in Figs.~\ref{DW_Motion_Example1} and \ref{DW_Motion_Example2}.
Each current pulse induces some damped oscillations of the DW near $x=0$. These are correlated with the rotation and movement of the DW orientation.
For short pulses with weak damping the polarization is not changed, see Fig.~\ref{DW_Motion_Example1}.
If we increase the damping then we find the polarization is changed by the first pulse and then
changed back by the second pulse, see Fig.~\ref{DW_Motion_Example2}.
If we then increase the length of the pulse then there is no change in polarization again.
Further increasing the pulse strength will change the polarization again.

\begin{figure}
\includegraphics[width=0.95\columnwidth]{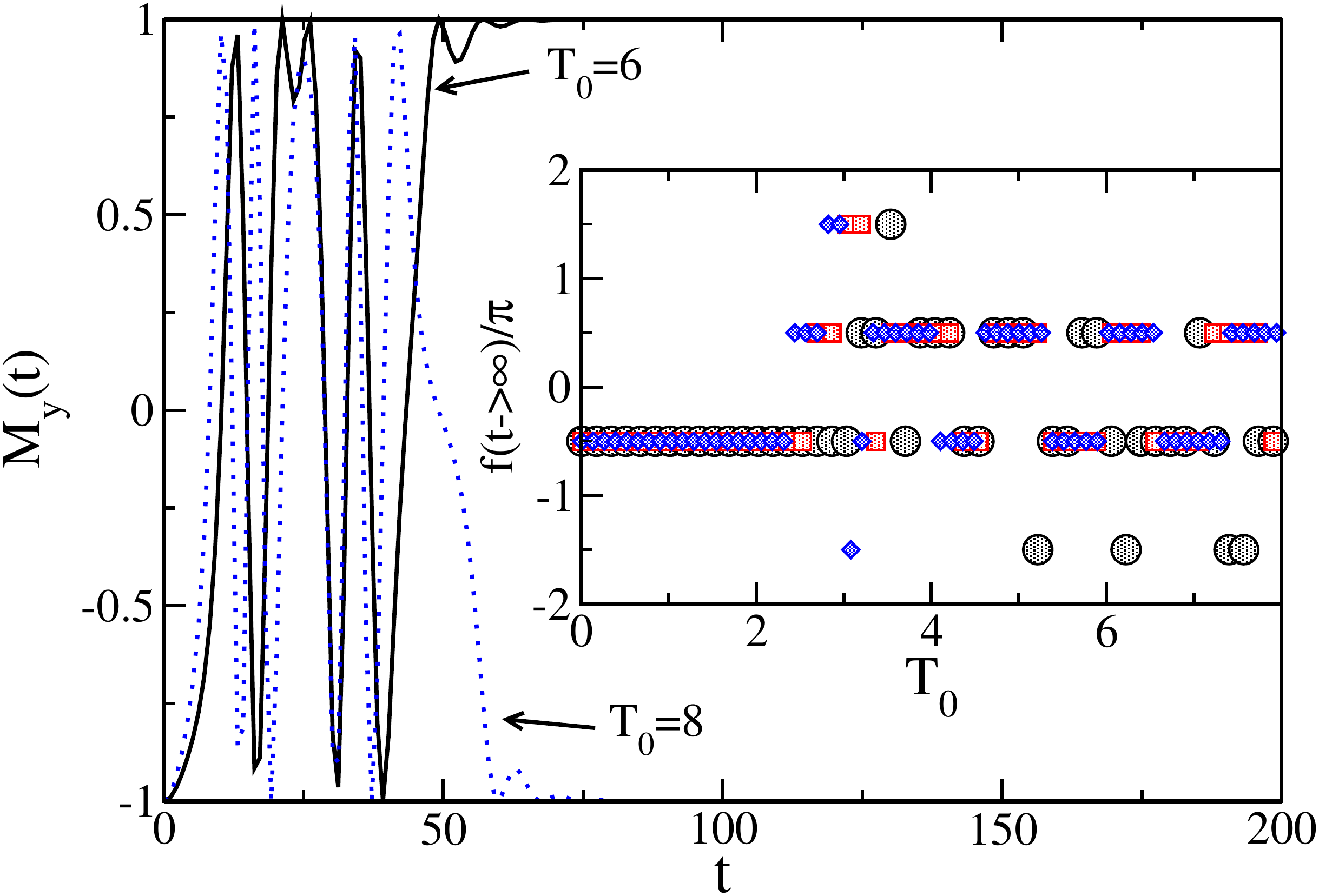}
\caption[DW Motion]{The DW motion for $T_t=\tilde{T}(t-20)$ with $\alpha=0.2$, see Eq.~\eqref{pulse}. The main figure shows $M_y(t)$ for different $T_0$ and $\beta=0.01$.
In the inset are results for $\beta=0.01$ (black circles), $\beta=0.008$ (red squares), and
$\beta=0.006$ (blue diamonds).}
\label{DW_Motion_Critical}
\end{figure}

As Eq.~\eqref{eqn_motion_f} depends also on the time derivative of the torque, we can effectively move
the DW with a smaller torque if it changes very quickly. To see this we plot the final DW position
for a variety of pulse strengths $T_0$ with different $\beta$.
As $\beta$ is decreased the necessary pulse strength to make a change in the polarization of the DW
is also decreased, see the inset of Fig.~\ref{DW_Motion_Critical}. Shown is the angle $f(t\to\infty)$,
i.e.~the final value of the DW polarization.
It should be noted that for increasing pulse strength this is not just a monotonically increasing function. All though we find a $\beta$-dependent minimum pulse strength which will rotate the DW, the final polarity of the    DW changes almost randomly as a function of increasing pulse strength, see Fig.~\ref{DW_Motion_Critical}.

To relate the results of our calculations to some material parameters,
one can choose for example $K_{\perp}= 10 ^{-4}$J/m${}^2$, corresponding to Co \cite{Johnson1996},
and $S = 10 ^{-16}$m${}^2$ as for a wire with a width of $10$nm.
This gives us the energy unit $E_0 = 10^{-20}$ J $\approx 60$meV.
In our numerical calculations we take $SK = 0.2 E_0$ which is
close to the parameters of Co or Fe. The other parameter, the pinning field, was taken to be $V_0\lambda = 2 E_0$.
The estimation of the DW width with these parameters is
$\lambda \approx \sqrt{J/K} \approx 10$nm.
The corresponding time unit, as used in Figs.~\ref{DW_Motion_Example1}-\ref{DW_Motion_Critical}, is then $\hbar/E_0\approx10^{-14}$ s$= 10$fs. Hence, the dynamics of interest here are in the picosecond regime.

\section{Conclusion}

We considered the dynamics of a strongly pinned DW in magnetic nanowire under short current pulses.
For this purpose we derived the equation of motion for the polarization parameter of the DW.
This equation includes longitudinal and transverse components of the spin torque.
The essential point is that the time derivative of the transverse torque also acts on the DW, which makes
it possible to enhance the effect by using rapidly changing, i.e. short, pulses. Our numerical calculations
allow us to visualize the dynamics when one changes the parameters of damping, pinning and
anisotropy. This fact points to the possibility of optimal control of DW motion in the spirit of Ref.\cite{prl99}.
For this purpose torque pulses, generated possibly with laser-induced current pulses, should be
 in the picosecond regime.

\begin{acknowledgement}
This work is supported by the National Science Center in Poland as a research project
in years 2011 -- 2014, by the DFG contract BE
2161/5-1, and by the Graduate School of MAINZ (MATCOR).
\end{acknowledgement}

\providecommand{\WileyBibTextsc}{}
\let\textsc\WileyBibTextsc
\providecommand{\othercit}{}
\providecommand{\jr}[1]{#1}
\providecommand{\etal}{~et~al.}

\end{document}